# Observation of impedance oscillations in single walled carbon nanotube bundles excited by high frequency signals

George Chimowa, Siphephile Ncube and Somnath Bhattacharyya[a]

Nano-Scale Transport Physics Laboratory, School of Physics, and DST/NRF Centre of Excellence in Strong materials, University of the Witwatersrand, Johannesburg, South Africa.

We report experimental observation of impedance oscillations in single-walled carbon nanotubes measured from 100 MHz to 65 GHz on coplanar wave guides and a power law dependence of the differential conductance with bias voltage. From the crossover of the real and imaginary parts of the complex impedance observed in the range of 10 GHz, we estimate a long lifetime of 15 ps that can support the claim of ballistic transport. By measuring the scattering parameters at high-frequencies of a few aligned single-walled bundles at low temperatures we show that, this observation is strongly influenced by the number of tubes available.

_________________________________________________________________

[a] Author to whom correspondence should be addressed. Electronic mail: Somnath.Bhattacharyya@wits.ac.za

Over the past few decades there have been both theoretical [1, 2] and experimental [3-5] efforts to understand one dimensional (1D) transport in DC and high frequency regimes. Calculations based on collective (electron) phenomena have predicted alternative current (AC) response that is chirality dependent in single-walled carbon nanotubes (SWNTs) even in the presence of contact effects and a universal inductive to capacitive transition in the imaginary admittance with increase in contact resistance [6]. On the other hand, a semi-classical model based on a transmission line by Burke *et al.* has predicted possible Luttinger liquid excitations in metallic CNTs achievable in the gigahertz regime by setting up standing voltage waves in the CNTs [7]. Some experimental RF measurements were done on individual or bundle of SWNTs, double-walled CNT (DWNTs) and multi-walled CNTs (MWNTs) but there has not been experimental evidence of some of the theoretical predictions [8,9]. This motivated us to probe the temperature dependent AC response and extend previous works on CNTs.

In the high frequency (HF) regime AC conductance is believed to be equivalent to a transmission line composed of kinetic inductance ($L_k$) and quantum capacitance ($C_q$) as additional parameters to the magnetic inductance and electrostatic capacitance [6, 8]. If the decay length (of voltage waves) is comparable to the mean free path of the carriers and the intrinsic resistance of material is in the linear response regime, the real and imaginary parts of the impedance will be characterised by an oscillatory behaviour corresponding to resonant peaks of the fundamental waves set by the finite length of the transmission line [6]. Experimentally there were several attempts to collectively excite electrons in CNTs and observe impedance oscillations using high frequency which were unsuccessful due to a couple of reasons such as (i) the weak CNT signal from individual SWNTs making it difficult to resolve from instrument noise or (ii) too many CNTs resulted in average (bulk-like) properties instead of the expected quantum effects. It is therefore important that a trade-off point between signal strength and number of CNTs be established without losing the individual properties. Another possible solution is to use bundles instead of single tubes. An immediate notable disadvantage of bundles is that many inter-tube entanglement which may result in defects.

Some DC experiments have shown that transport in bundles is dominated by individual metallic tubes [9-11], and if these are weakly coupled observation of impedance oscillations should still be possible depending on the amount of defects in the SWNT bundles. We thus succeeded in this by firstly reducing the material dimension so that it becomes comparable to the electron mean free path and secondly using time dependent signals, by increasing the frequency of the stimulus signal such that it is higher than the scattering rate in the material. This was an attempt to get around the problem of defects that are inevitable in chemically purified and lithographically contacted CNTs. We therefore report direct observation of the oscillatory impedance in aligned bundles of SWNT using HF excitations by performing detailed measurements of transport properties over a wide range of frequency and temperature. In previous works on high frequency properties of dense DWNTs networks, fabricated by ink-jet printing onto the waveguide, a strong frequency dependence of DWNTs microwave parameters as well as abrupt changes in effective permittivity was observed [12]. This property could be used in gas sensor applications. We extend on this work by probing few aligned tubes so as to unearth any unique quantum features regarding this strong frequency dependence.

Single walled carbon nanotubes SWNTs were synthesized by laser ablation and purified by chemical means, see ref [13] for further details. They were then dispersed in dichlorobenzene and prepared for S- parameter measurements by aligning them on 1000 µm long coplanar waveguides (CPW) with a 1 µm gap using dielectrophoresis. A set of four nano-manipulators (from Kleindiek) mounted in the SEM were used to remove unwanted CNTs. Small contacts were made using a combination of e-beam lithography (RAITH – Germany) and a gas injection system from Omni-GIS. The schematic diagram in Fig 1(a) – (d) illustrates the whole device fabrication procedure. The waveguides are made of a bilayer of titanium and gold. Some of actual SEM micrographs of the devices that were used for HF measurement are shown in figures 2(b-c).

S- Parameter measurements were done using a cryogenic micro-manipulated probe station from Janis and an Agilent Precision Network Analyser (PNA) model E8361C at room temperature under vacuum ($10^{-5}$ mbar) in a dark cryostat to try and eliminate photon assisted tunnelling which tends to complicate the analysis. The stimulus power was set at -17 dBm for both probes which is equivalent to 31.6 mV so that the ac current is below saturation and ensure we promote low energy excitations. The measurements were done from 0.01 to 65 GHz and the IF bandwidth was reduced to 500 Hz to improve the accuracy by reducing random noise from the PNA. PNA calibration was done using the short-open-load-through (SOLT) of the GGB CS – 15 calibration kit to set the reference of the signal at the probe tips. The *I-V* measurements were done using the Agilent B1500A semiconductor analyser with a 1 pA resolution. Measurement parasitics were extracted by using the open – short de-embedding method an industrially accepted technique, see equation one [14].

$$Y_{DUT} = [(Y_{msd} - Y_{open})^{-1} - (Y_{short} - Y_{open})^{-1}]^{-1}. \qquad (1)$$

In this matrix (inverse) equation $Y_{msd}$ is the measured admittance of the CPW plus the aligned CNT, $Y_{open}$ and $Y_{short}$ are the admittances of the open (CPW without CNT) and short (CPW without a gap) dummies. The device under test (DUT) i.e. CNTs data can then be obtained by converting the admittance (*Y*) to *S*- or impedance (*Z*) data using standard conversion equations [15].

(i) DC Conductance measurements: As part of device quality check, we measured the *I-V* characteristics and it was observed that all the devices did not show any current saturation up to 2 V. Furthermore the current (*I*) was found to be proportional to $V^{1+\varphi}$, with φ ranging from 0.42 to 1.42 for different devices (see inset of Fig 2(a)). This is similar to other reports given in references [16-18]. The main graph (Fig 2(a)) further show a collapse of various temperature curves which is similar to our previous report [14]. The graphs shows that the condition is valid above 1 V satisfying the condition *eV* >> $k_\mathrm{B}T$.

From the power law model, the exponent ($\varphi$) of $V$ depends on the conduction channels as well as on how the electron tunnels into the excited system of electrons in the CNT. It therefore gives us an indication of the correlation of the system if we take note of the exponent of the power law which is less than one, indicating a strongly correlating electron system.

(ii) Room temperature AC Impedance measurements: Scattering parameter measurements are very sensitive to temperature, calibration, instrument drift and the contact force exerted by the HF probe onto the CPW. It is therefore paramount that these factors are handled with care as they may introduce large systematic and random errors. From a number of systematic checks we observed that the highest possible error is less than 0.4 % of a dB in the transmitted scattering parameters. From repeated measurements we have shown that our results are reproducible within the mentioned accuracy, secondly there is insignificant instrument drift and lastly that the solvent used to disperse the CNTs does not influence the measurements [9].

Figure 3, shows the measured transmission scattering parameters ($S$12), for a selected set of samples, SWNTs (SW2 – with 6 CNT bundles, SW7 – with 2 CNT bundles) and that of the open and short dummy waveguide. The numbers on the labelling are a representation of the position of the specific device on the waveguide substrate. The data for the dummies is used to remove parasitic effects, a process known as de-embedding. It is evident from the graph (and Inset) that the average power transmitted by CNTs only is on average about 3 dB obtained from $\Delta S_{CNT} = S_{DUT} - S_{open}$. The measurement uncertainty is about 0.06 dB at high frequencies (~ 50 GHz) and this would mean a signal to noise ratio of about 50 dB. After removing the waveguide parasitic effects using the open-short method [20], we notice a clear distinction in the real and imaginary components of the CNT's impedance shown in figure 4(a) and 4(b).

A clear outstanding feature from the graph (Fig. 4(a)) is the oscillatory impedance with an almost constant period from 1 GHz to about 48 GHz. The oscillation amplitude for Real Z gets exponentially damped as the frequency increases probably due to weak energy dissipation from the waveguide. The graph also show increase in transmission in the CNTs with frequency as the impedance decreases into the Ohms range at high frequency. The imaginary component impedance also shows similar oscillatory behaviour and a change from capacitive to inductive nature as the frequency increases see Fig. 4(b). A similar observation was made for other sets of samples which are not shown here.

The reported nontrivial frequency dependence (resonance) in the impedance should be expected if the electron mean free path is comparable or greater than the sample dimension according to transmission line theory [21]. Since the vector network analyser stimulus was set at -17 dBm (equivalent to 31.6 mV), the oscillations should be due low lying energy excitations and thus should be able to give insight into the correlation of electrons. The HF measurements done at low temperatures show lowering the temperature (i.e. suppressing phonon effects) results in the impedance oscillations being washed out [see Fig 5(a)]. The immediate explanation is that at low temperatures the effective electron wave velocity is increased implying that the oscillations should be observed in the sub-THz regime because of the dimension of the transmission line.

Low temperature measurements should unveil ballistic transport which is regarded as the motion of charge carriers without any scattering. However contact resistance always adds a finite resistance to the system. We have attempted to achieve ballistic transport in two ways; (i) reducing the sample dimensions to the range of the mean free path or (ii) using a time-dependent signal whose frequency is greater than the scattering rate in the material. In this case the effect of local defect centres to transport can be reduced. The Drude model for AC transport expresses the conductivity made of two parts; the real part and imaginary parts given by $\sigma(\omega) = \frac{ne^2\tau}{m^*} \frac{1}{1+i\omega\tau}$, where $\tau$ is the momentum scattering time, $m^*$ is the effective electron mass, and $n$ is the carrier density. The imaginary part of the conductivity is given by $i\omega\tau$ and it changes sign depending on whether it is capacitive or inductive. We can therefore make use of this fact to determine experimentally whether the transport is the diffusive or ballistic regime if the real and imaginary impedance is known as shall be explained in the results section.

At room temperature when phonon scattering is significant within the CNTs, the ballistic regime can still be realised when the stimulus frequency is greater than the scattering rate in the CNTs. This is shown in Fig (5(b)), which shows a transition from diffusive to ballistic conduction at approximately 13 GHz in SW7. This happens when the real and imaginary impedances are equal and the contact resistance is neglected (which is a challenge at the moment to achieve experimentally). At this crossover point $\omega\tau = 1$, where $\omega$ is the frequency and $\tau$ is the momentum scattering time. It therefore means the momentum scattering time in the SWNTs will be approximately 15 ps, which implies scattering length several times greater than sample length and resonance quality ($Q$) factor greater than one. However since we had finite contact resistance it means this crossover is happening at slightly lower frequencies than what the graphs actually suggest.

It is also evident in Figure 4(a) that the resonant peaks in the real impedance occur at the time the imaginary component changes sign, a clear indication of phase shifts during resonance. The period of oscillations was 63 ps, the period is about 90 ps for device SW7. The only significant difference is in the damping of the amplitude of oscillations in the experimental data. This can be explained in terms of the damping due to the intrinsic resistance of the CNTs or energy losses in the transmission line of which the theoretical model does not take into account. A very similar kind of impedance oscillation was explained theoretically in previous report based on the claim of Luttinger liquid [7,20]. However, to establish the origin of these oscillations based on collective modes much more work is needed.

In an attempt to determine the magnitude of the capacitance and inductance values for our devices we simulated the transmission coefficient experimental data (S21) with an equivalent circuit (using MATLAB RF Simulink). Figure 6 shows the fitted S21 data and the inset is the equivalent circuit for one of the device SW7. The obtained circuit parameters (R ~ 1.3 kΩ/μm, L ~ 7.4 nH/μm and C ~ 66 fF/µm were obtained after subtracting the effects from the open dummy and they are found to be in the order of magnitude for four CNTs in parallel. It is also based on the assumption that the kinetic inductance is considered to be the dominate inductance and the capacitance is the total of the electrostatic and quantum components because in HF measurements the two play a significant role [13]. We attribute these observations of the elusive quantum effects to the few numbers of CNTs bundles aligned between the CPW and the design of the CPW such that the capacitance of the signal gap does not short (circuit) the CNTs at high frequency.

In view of the above discussions we have shown experimentally features that are associated with the impedance oscillations in CNTs using high frequency stimulus in the GHz range. These oscillations are from the resonance of voltage waves set in the CNT by their finite length. We have also shown a crossover from diffusive to ballistic transport achieved by increasing the stimulus frequency to higher than the scattering rate. This work has shown that few CNT conduction bundles are needed to observe the elusive quantum features and measurable quantum parameters such as capacitance and inductance. These are measurable parameters can be incorporated in circuit design software's for electronic applications. High frequency studies of this nature will lay a new foundation in understanding and making of a new class of fast switching molecular carbon based electronic devices with enumerable advantages.

*Acknowledgements:* SB acknowledges the CSIR-NLC for establishing the laser ablation set up which produced the SWNT samples and NRF (SA) NNEP as well as nanotechnology flagship grant.

**Figure captions:**

**Fig 1:** Schematic diagram of the whole fabrication process of the devices and the measurement setup. It consists of CNT alignment using dielectrophoresis **(a)**, CNT selection using Nano-manipulators **(b)** and CNT metal contact bonding using a combination of e-beam lithography and gas injection system (GIS) **(c)**. The HF impedance is then measured using G-S-G probes with a 50 um pitch size **(d)**.

**Fig 2(a):** The main figure shows that the differential conductance for SW7 at different low temperatures collapsing to a single gradient at high bias voltages. The inset shows the power law dependence of the current at 29.5 K for SW2. **Fig 2(b & c):** Shows SEM micrographs of SW2 and SW7 devices, respectively.

**Fig 3:** S-parameter ($S_{12}$) data for the two CNT devices SW2 and SW7, which had six and two SWNTs respectively. The short and open data is for dummies used for de-embedding. The inset is a magnified version of the same data at high frequencies above 40 GHz.

**Fig 4(a): S**hows the Real impedances for samples SW2 (with 6 SWNT bundles) and SW7 (with 2 SWNT). **(b)** Shows the imaginary component of impedances for the same samples (SWNT). All the figures are characterised by oscillatory impedance which not present for the open dummy.

**Fig. 5(a):** Low temperature impedance for SW2 showing washed out oscillations. **(b)**The graph shows a crossover of the Real and Imaginary impedances of SW7 (with 2 SWNT bundles) this is a sign of a crossover from diffusive to ballistic transport as the stimulus frequency becomes greater than the scattering rate.

**Fig. 6:** Simulated and experimental transmission coefficient data for SW7 device. The inset is the equivalent circuit used to simulate the data.

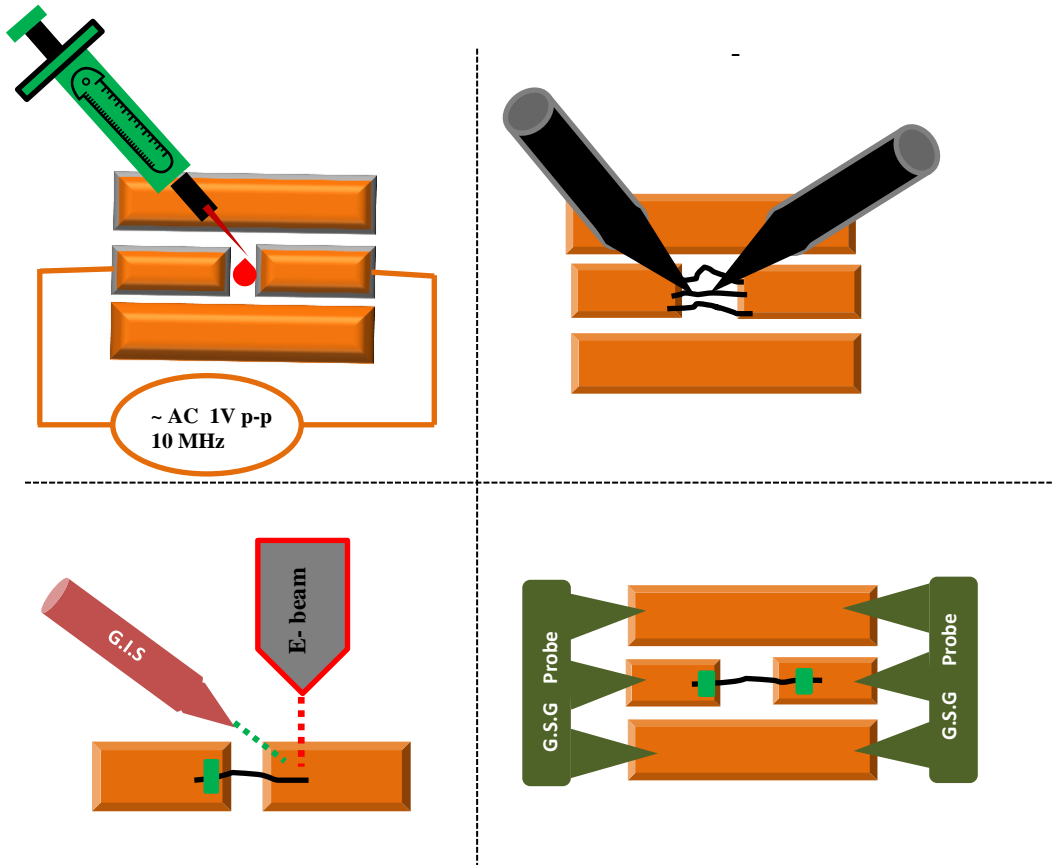

**Fig.1:** Chimowa *et al*.

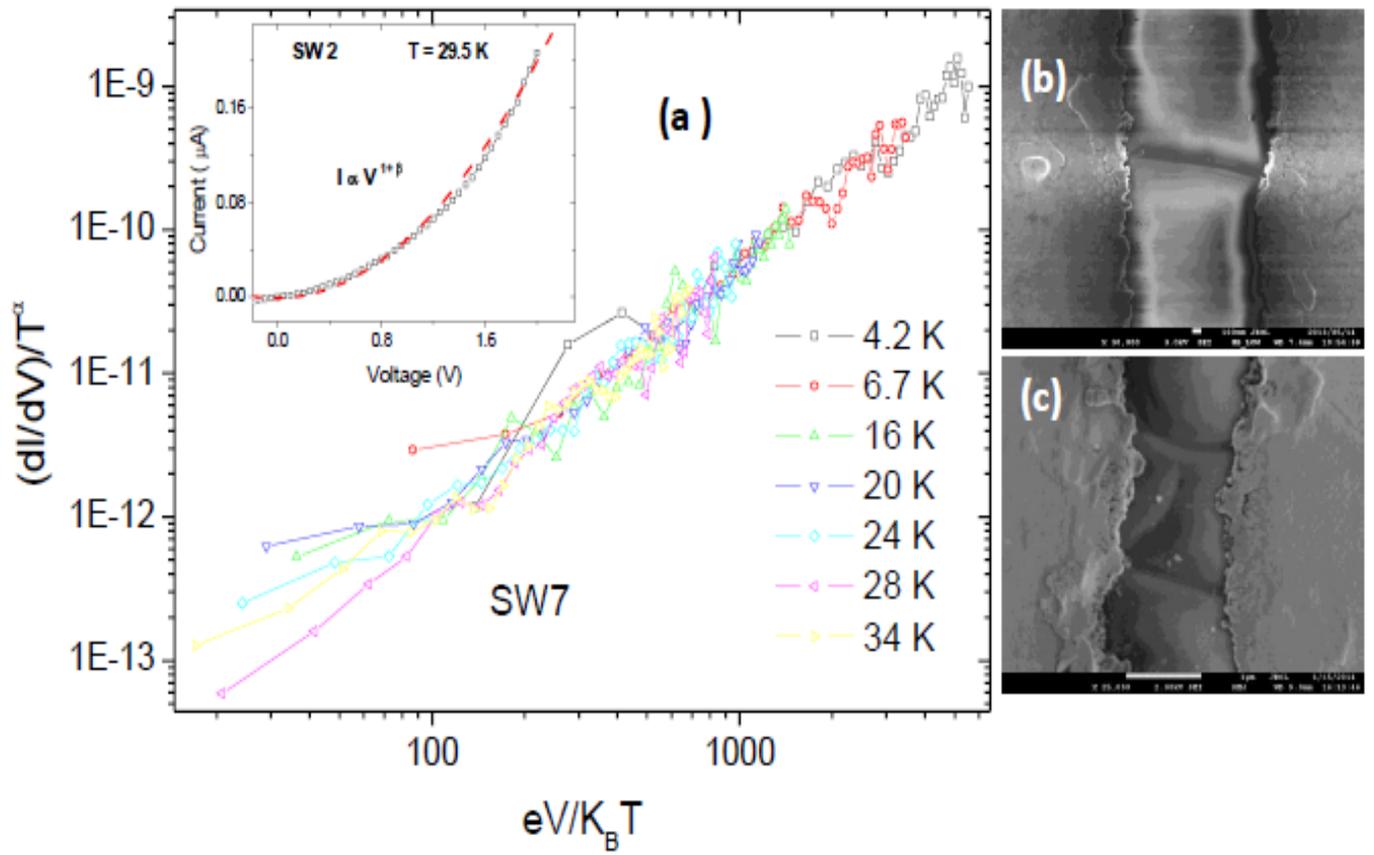

**Fig. 2 (a,b & c):** Chimowa *et al*.

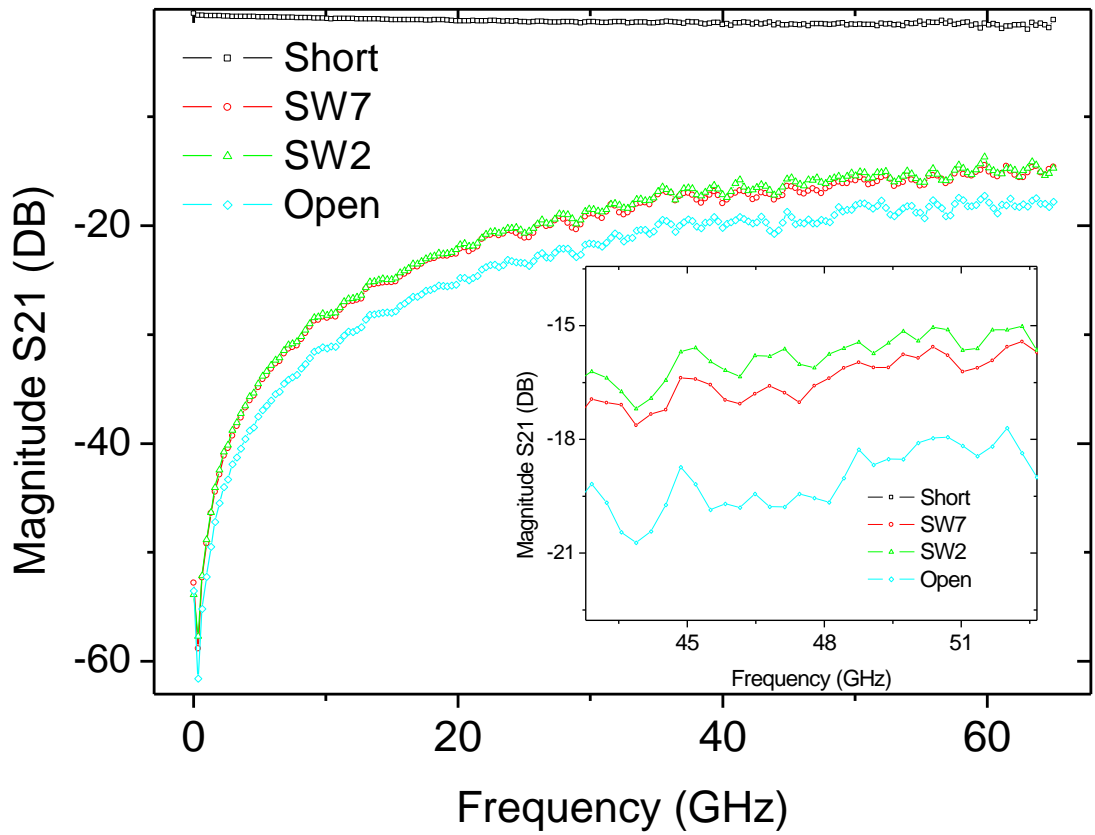

**Fig. 3:** Chimowa *et al*.

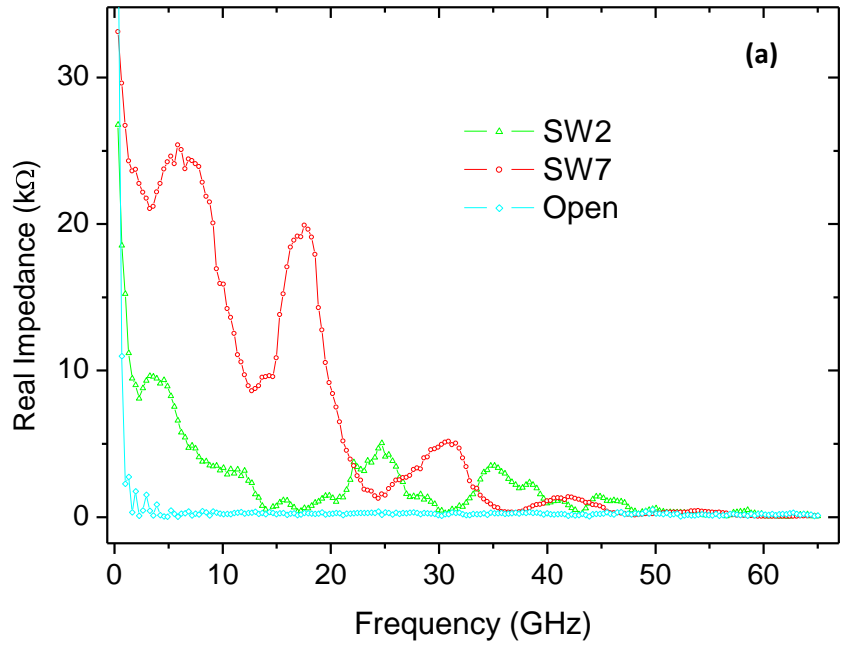

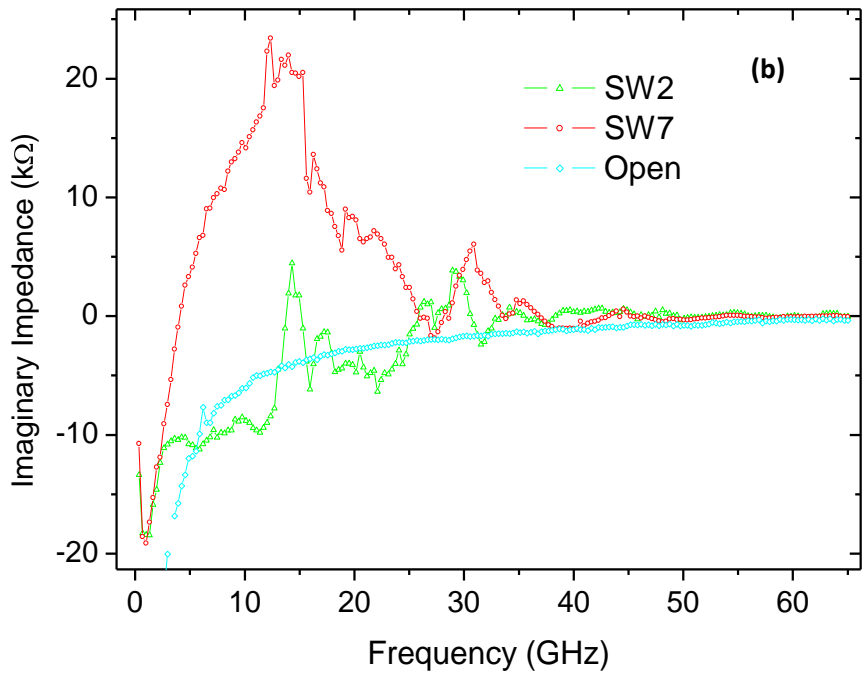

**Fig. 4(a) and (b)**                                                                 Chimowa *et al*.

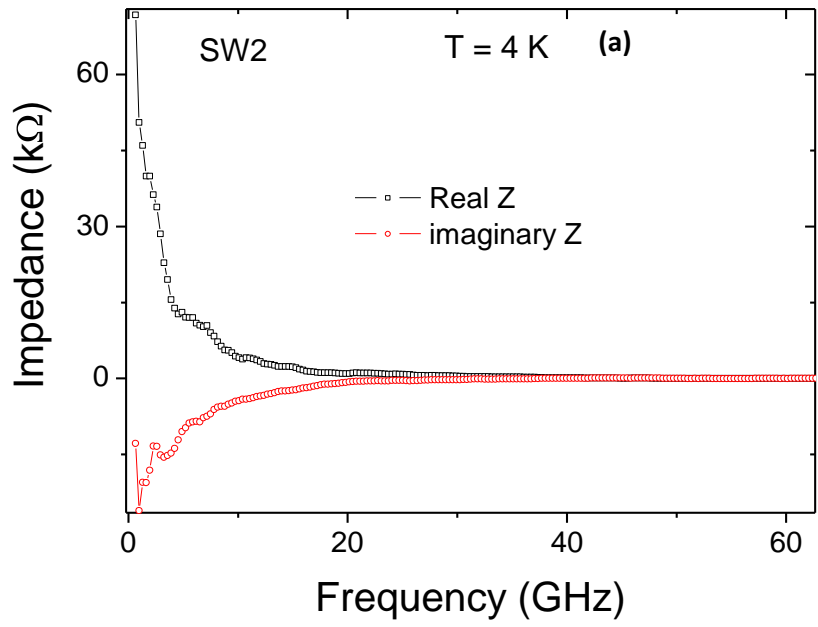

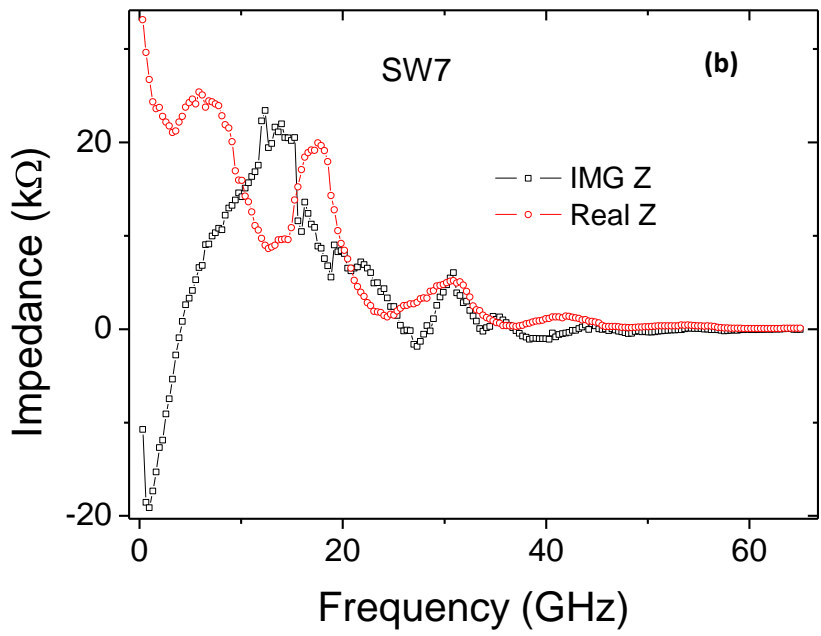

**Fig. 5(a) and (b)**  Chimowa *et al*.

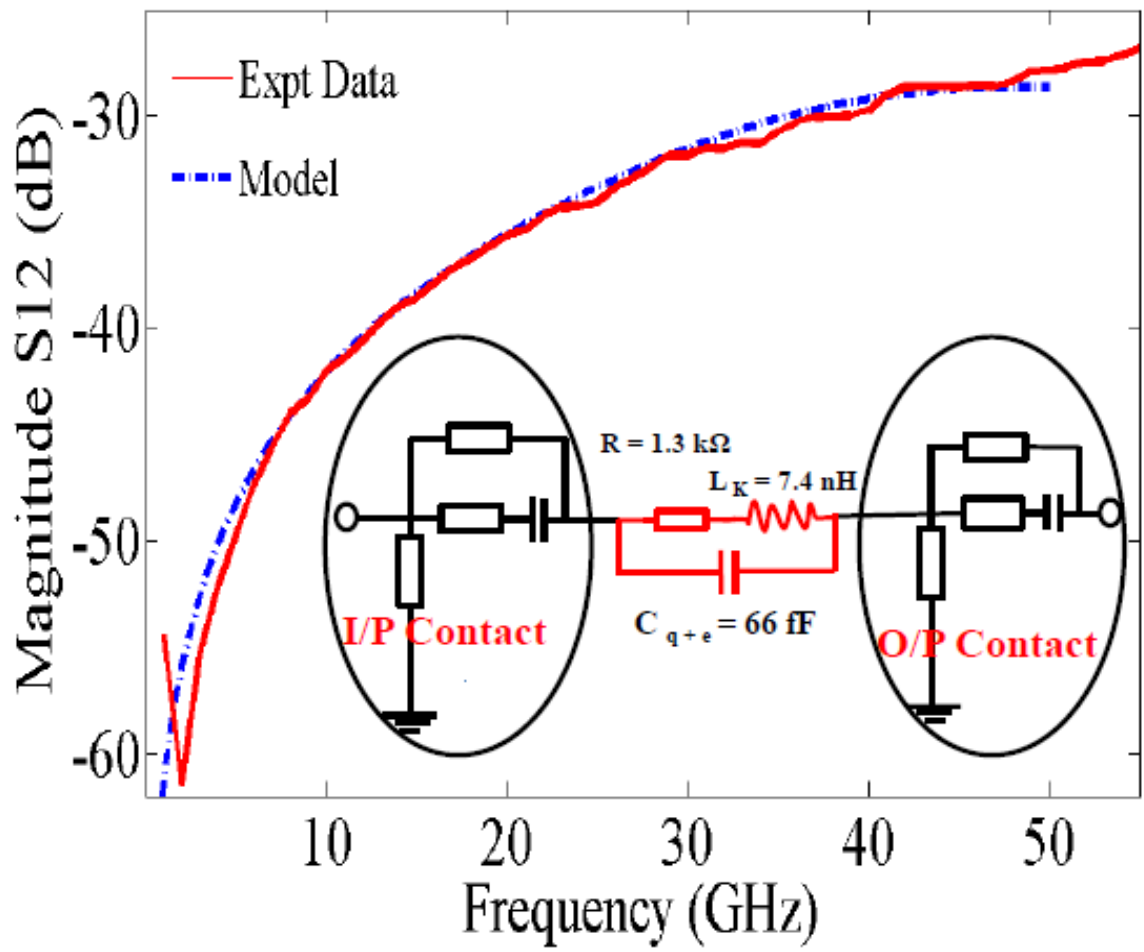

**Fig. 6.**   Chimowa *et al*.